\documentclass[lettersize,journal,10pt,comsoc]{IEEEtran}
\pdfoutput=1 
\usepackage{amsmath,amsfonts}
\usepackage{epstopdf}
\usepackage{algorithmic}
\usepackage{algorithm}
\usepackage{array}
\usepackage[caption=false,font=normalsize,labelfont=sf,textfont=sf]{subfig}
\usepackage{textcomp}
\usepackage[hidelinks]{hyperref}
\usepackage{stfloats}
\usepackage{url}
\usepackage{verbatim}
\usepackage{graphicx}
\usepackage{cite}
\usepackage{setspace}
\usepackage{booktabs}
\begin{document}

\title{Interference-Managed Local Service Insertion for 5G Broadcast}

\author{M. V. Abhay Mohan\textsuperscript{\href{https://orcid.org/0000-0002-9290-9929}{\includegraphics[scale=.04]{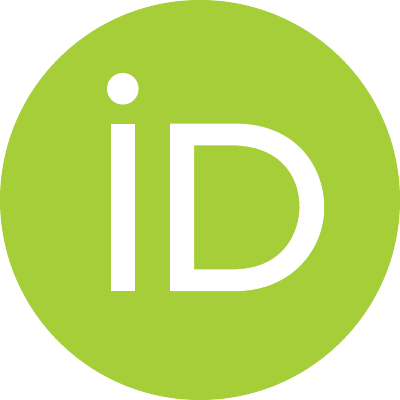}}}, and K. Giridhar\textsuperscript{\href{https://orcid.org/0000-0001-6044-2036}{\includegraphics[scale=.04]{ORCIDiD_iconvector.pdf}}}
\thanks{Manuscript received XXXX, 20XX; revised XXXX, 20XX.}%
 \thanks{The authors are with the Department of Electrical Engineering, Indian Institute of Technology Madras, Chennai, Tamil Nadu, India (e-mail: abhay@telwise-research.com, giri@ee.iitm.ac.in)}
 }%

\markboth{}%
{M. V. Abhay Mohan and K. Giridhar: Interference-Managed Local Service Insertion for 5G Broadcast}


\maketitle

\begin{abstract}
Broadcast of localized TV content enables tailored content delivery catering to the requirements of regional user base. 5G multicast-broadcast service (MBS) requires a spectrally efficient broadcast solution that enables the change of content from one local service area (LSA) to another. A frequency reuse factor of unity between two adjacent LSAs causes their boundary region to become saturated with co-channel interference (CCI). Increasing the reuse factor will reduce the CCI at the cost of degrading the spectral efficiency. This paper addresses the frequency and transmit power planning which manages the CCI at the LSA boundary to achieve a satisfactory trade-off between spectral efficiency and broadcast coverage.
\end{abstract}


\section{Introduction}
\IEEEPARstart{C}{ellular} broadcast enables resource-efficient multimedia content delivery to mobile users. The broadcasting service in 5G-NR is required to support ``local, regional and national broadcast areas" \cite{etsi_5G}.  Content that is of interest to a wide area, like an entire nation, is called global content. Content that is broadcast to a smaller region is called local content. Broadcast uses single frequency networks (SFNs) which yields diversity advantages from synchronized transmissions of the same content from neighboring towers. However, co-channel interference significantly limits reception when two distinct local content transmissions in adjacent local service areas (LSAs) use the same time-frequency resources (reuse-1) \cite{EBU027,WeiLi2019Coverage}.  

Conventional terrestrial broadcast systems such as the advanced television systems committee (ATSC) and digital video broadcasting-terrestrial (DVB-T) cover large geographical areas of the order of hundreds of kilometers using high-power, high-tower (HPHT) transmitters. Local service insertion is achieved using layer division multiplexing (LDM) \cite{Li2017Using} in ATSC 3.0 systems. This scheme transmits a highly robust core layer (CL) signal with a higher power over a less robust low-power enhanced layer (EL) signal. Both CL and EL occupy the same bandwidth and can be sequentially decoded with the help of successive interference cancellation. However, local content served with the EL can reach the cell edge only with the help of a rooftop directional antenna \cite{WeiLi2019Coverage}. Moreover, local content will not receive the diversity advantage of SFN that is typical in broadcasting systems, as its transmission barely reaches the cell edge.

The DVB-next generation handheld (NGH) standard uses hierarchical local service insertion (H-LSI) to transmit a low-priority local service on top of the high-priority global service. In this scheme, local content can be received only in areas surrounding the transmitters \cite{Lopez2014Technical}. Alternately, this standard also supports orthogonal local service insertion (O-LSI). Nearby LSAs that transmit distinct content are allotted non-overlapping orthogonal frequency division multiplexing (OFDM) subcarriers in O-LSI. Other ways of achieving orthogonality, such as placing interfering contents in distinct carriers or subframes, are described in \cite{EBU027}.

For cellular broadcast, techniques such as LDM and H-LSI cannot give good local content coverage as they perform best with a fixed rooftop receiver. The cellular receiver may be indoors, inside a vehicle, or somewhere without a direct line-of-sight path. In such situations, the signal-to-noise ratio (SNR) reduces considerably, and the local content transmission using LDM or H-LSI would be almost impossible to decode. Orthogonality-based approaches such as non-overlapping subcarrier sets as in O-LSI, using distinct carriers or subframes as in \cite{EBU027}, etc., will enable the local content to reach the cell edge. In addition, nearby towers that transmit the same local content can be synchronized to provide an SFN diversity gain.  

However, the orthogonal broadcast techniques have low spectral efficiency as parts of the spectrum must be globally reserved for different local contents. The local content will have a frequency reuse factor of 2 or more, while the global content will have a reuse factor of 1. This avoids co-channel interference (CCI) at the boundary between different LSAs. This paper discusses two schemes that can be used at the LSA boundary to improve the overall spectral efficiency compared to techniques like O-LSI. While one of the proposed schemes scales down the power of the local content near the LSA boundary, the other scheme restricts O-LSI-based local content broadcast to the boundary cells alone. As both techniques advocate interference management rather than avoidance, we call them interference-managed local service insertion (IM-LSI) in this paper. The proposed approaches have higher spectral efficiency and enable SFN diversity gain for the local contents. An earlier draft of this work is available as a preprint in \cite{mohan2022novel}, however it does not include the proposed restricted O-LSI approach mentioned above.

\section{System Model}
\begin{figure}
\centering
\includegraphics[width=0.8\columnwidth]{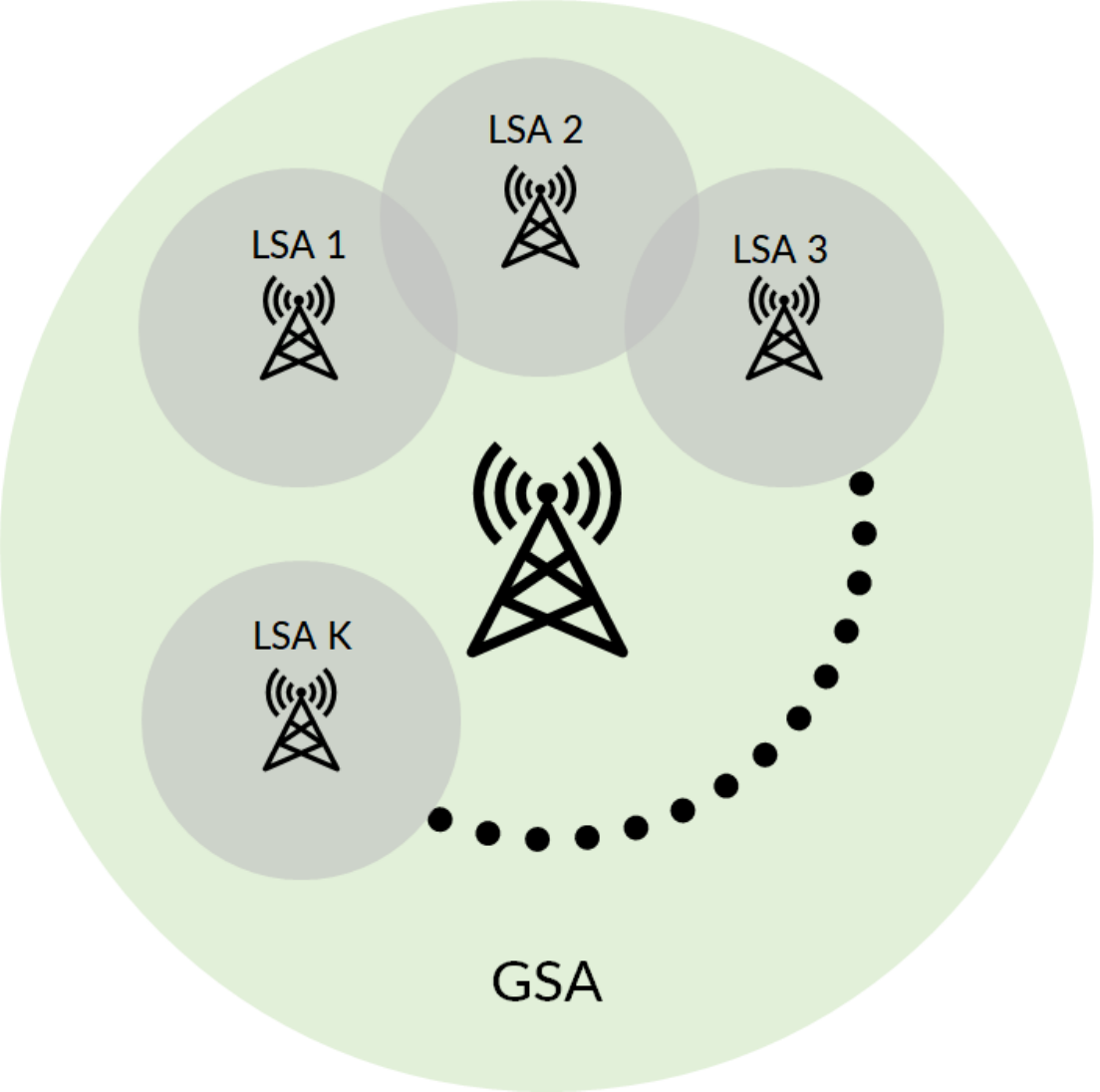}
\caption{System model for $K$ LSAs within a global service area}
\label{GSAandLSAs}
\end{figure}

Fig. \ref{GSAandLSAs} shows a cellular broadcast system consisting of $K$ LSAs within a global service area (GSA). Each tower in the figure is representative of all the transmitters required for the SFN in that particular service area. Although the figure shows a specific LSA overlapping with two other LSAs, all K LSAs can overlap with each other in general. 

Fig. \ref{CellGrid} shows the constituent cells in two adjacent LSAs. Each cell is assumed to have an omnidirectional transmitter. The cell grid consists of $8$ rows and $10$ columns of cells. The cells are divided evenly, with $40$ cells in LSA1 and $40$ in LSA2. The boundary region between the LSAs shall be called the buffer region in this work. The buffer region consists of the left buffer (marked LB) which belongs to LSA1, and the right buffer (RB) which belongs to LSA2. The LSAs serve distinct local contents in the same set of subcarriers. This causes CCI for the local contents at the buffer zone. The simulation studies presented in this work consider either area A1 which exclusively considers LSA1, or area A2 which covers both LSAs.


The broadcast information from each tower is OFDM modulated before transmission. If a broadcast system with $M$ contents is assumed, the signal powers of contents in LSA1 are denoted as $S_1, S_2, ... S_{M}$, with the index $1$ always indicating the global content. For LSA2, the signal powers are $S_1, S_2', ... S_{M}'$. The global content occupies the subcarriers in set $\mathcal{S}_1$. The local contents in LSA1 occupy non-intersecting sets of subcarriers $\mathcal{S}_2, ... \mathcal{S}_{M}$ while those in LSA2 occupy  $\mathcal{S}_2', ... \mathcal{S}_{M}'$. The cells that transmit a particular content are indexed by $l \in \Gamma_{\mathcal{S}_m}$, where $\Gamma_{\mathcal{S}_m}$ is the set of all cell indices which transmit the $m^{th}$ content of LSA1 that lies outside the buffer zone. Here, $\Gamma_{\mathcal{S}_m'}$ is the set of all cell indices which transmit the $m^{th}$  content of LSA2 which lie outside the buffer zone. The sets of cells in these LSAs that lie in the buffer zone, LB and RB, are denoted by $\Gamma_{\mathcal{\tilde{S}}_m}$ and $\Gamma_{\mathcal{\tilde{S}'}_m}$, respectively. Since all the towers transmit the global content $m=1$, $\Gamma_{\mathcal{S}_1 \cup \mathcal{\tilde{S}}_1 }$ is the index set of all cells in LSA1.

\section{SINR and Spectral Efficiency}
\subsection{Orthogonal Local Service Insertion}
\subsubsection*{(A1) SINR of O-LSI}
We consider the scenario where there is sufficient spectrum availability to broadcast $M-1$ local contents in non-intersecting sets of subcarriers. Assume that invariably only two LSAs are expected to interfere with each other in a particular geographical region. In the O-LSI model, one local service area may be given $\lfloor \frac{M-1}{2} \rfloor$ local services while the other may be allotted $\lceil \frac{M-1}{2} \rceil$ local services. This design avoids CCI, and the signal-to-interference-plus-noise ratio (SINR) reduces to SNR. However, the number of local contents available per LSA is halved. We denote the SINR for the $m^{th}$ local content at coordinates $(x_0,y_0)$ as $SINR(m,x_0,y_0)$ as $\gamma_m$, given by

\begin{equation}
\gamma_m= \dfrac{\sum\limits_{l\in \Gamma_{\mathcal{S}^{(O)}_m \cup \mathcal{\tilde{S}}^{(O)}_m}} \dfrac{S_m}{ {d^{\ \eta}_{l,(x_0,y_0)}} }}{N_0 B_m}
\label{OLSI_SNR}
\end{equation}
The notation $^{(O)}$ makes it explicit that the O-LSI scheme is used. Here, $d_{l,(x_0,y_0)}$ is the Euclidean distance from the tower in cell $l$ to point $(x_0,y_0)$, $\eta$ is the path loss exponent, $N_0$ is the noise spectral density, and $B_m$ is the bandwidth of the $m^{th}$ content. This SINR expression assumes that the guard interval is long enough to accommodate the multipath components from the various towers in the SFN. If this is not the case, a weighing function similar to the one proposed in \cite{Rong2008Analytical} should be employed. The effect of fading and shadowing can also be modeled into \eqref{OLSI_SNR}, but it is assumed here that these effects have been averaged out. The assumptions for the SINR expression used in this work are similar to that used in \cite{liu2020towards}.

\begin{figure}
\centering
\includegraphics[width=0.8\columnwidth]{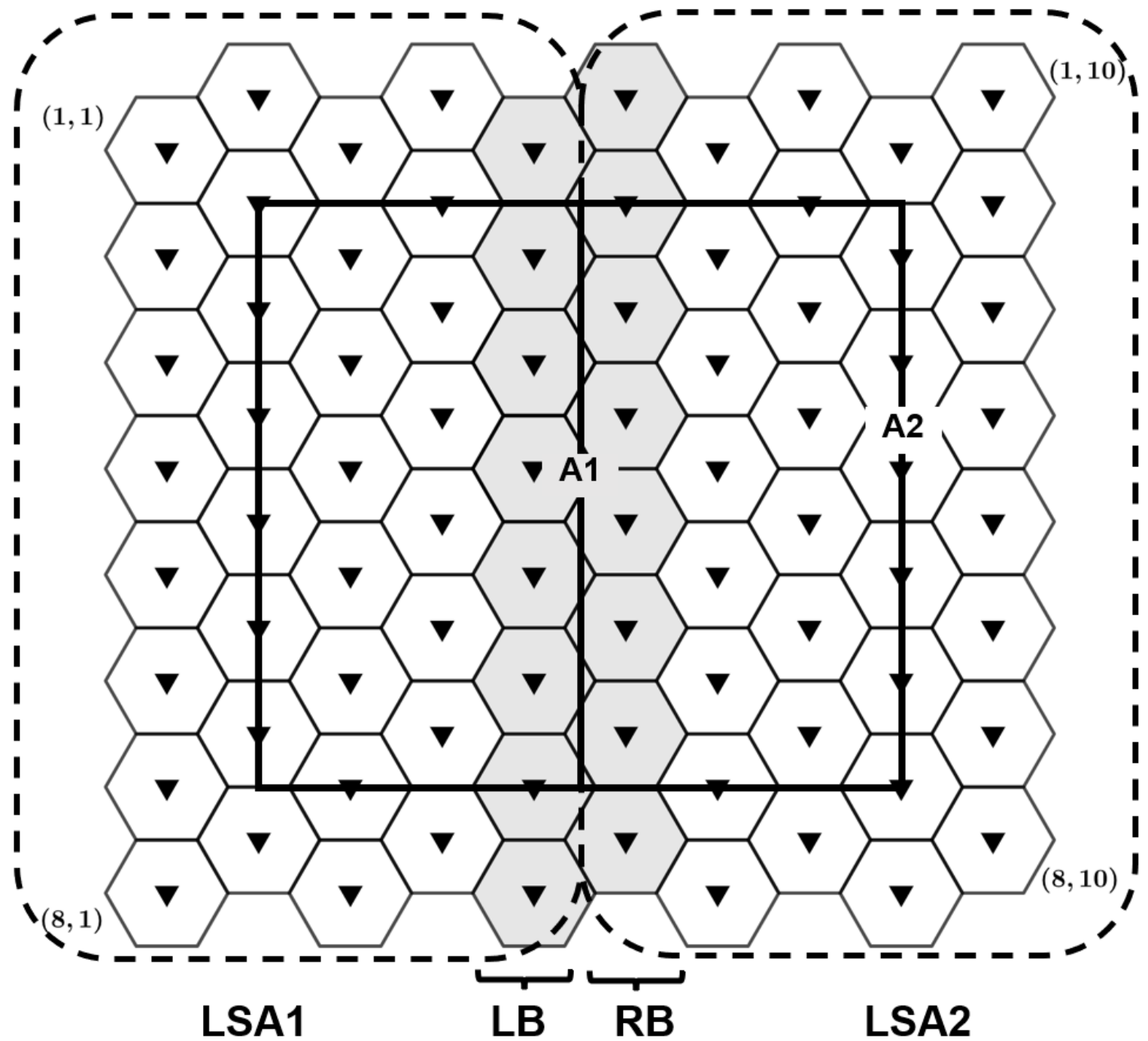}
\caption{Two LSAs and the buffer region at their boundary}
\label{CellGrid}
\end{figure}
\subsubsection*{(A2) Spectral efficiency of O-LSI}
The theoretical peak spectral efficiency of O-LSI can be calculated as follows. The product of the number of subcarriers allotted to a particular content and the number of bits per subcarrier is divided by OFDM symbol duration to obtain the theoretical bit rate. In O-LSI, half of the local contents are not transmitted in a particular LSA. This has to be accounted for in the expression. The theoretical bit rate is thus weighed by the fraction of cells in the LSA that transmit the $m^{th}$ content. This fraction is either $0$ or $1$ for O-LSI. The weighed theoretical bit rate is summed over all contents and divided by the total bandwidth allotted to broadcast transmission to obtain the overall spectral efficiency. 
\begin{equation}
    \xi_{O-LSI} = \frac{\sum_{m=1}^M \left( \frac{|\Gamma_{\mathcal{S}^{(O)}_m \cup \mathcal{\tilde{S}}^{(O)}_m }|}{|\Gamma_{\mathcal{S}^{(O)}_1 \cup \mathcal{\tilde{S}}^{(O)}_1}|} \frac{|\mathcal{S}_{m}|\times log_2 \mu_m}{T_{sym}}\right)}{\sum_{m=1}^M B_m} \text{ bits/s/Hz}
    \label{SEOLSI}
\end{equation}
Here $\mu_m$ is the modulation order of the $m^{th}$ content. $|\mathcal{S}_{m}|$, the cardinality of the set $\mathcal{S}_{m}$, denotes the number of subcarriers allotted to the $m^{th}$ local content. $T_{sym}$ is the OFDM symbol duration. The total number of cells in LSA1 that transmit the $m^{th}$ content is given by $|\Gamma_{\mathcal{S}^{(O)}_m \cup \mathcal{\tilde{S}}^{(O)}_m }|$, while the total number of cells in LSA1 is given by $|\Gamma_{\mathcal{S}^{(O)}_1 \cup \mathcal{\tilde{S}}^{(O)}_1 }|$.

\subsection{IM-LSI through power scaling}
Our proposed IM-LSI method allows reuse-1 of local contents by either power or frequency allotment adjustments in the buffer zone. IM-LSI through power scaling improves local content coverage by taking advantage of the fact that for fixed signal power, the SINR improves with a reduction of interference power. 

\subsubsection*{(B1) SINR for power scaling approach}
In Fig. \ref{CellGrid}, the set of cells in RB are the closest interferers to LSA1. Thus, scaling down the transmission powers of these local contents reduces the dominant CCI to LSA1. The converse is also true for the transmission powers in LB and the CCI to LSA2. The ratio of the transmit power of a cell in the buffer zone to that in the SFN zone is denoted by $\beta$. The SINR expression for a local content ($m > 1$) is given by,
\begin{multline}
\gamma_m = \frac{\sum\limits_{l\in \Gamma_{\mathcal{S}^{(PS)}_m}} \frac{S_m}{ {d_{l,(x_0,y_0)}}^\eta } + \sum\limits_{l\in \Gamma_{\mathcal{\tilde{S}}^{(PS)}_m}} \frac{\beta S_m}{ {d_{l,(x_0,y_0)}}^\eta }}{ \sum\limits_{l'\in \Gamma_{\mathcal{S'}^{(PS)}_m}} \frac{S_m'}{ {d_{l',(x_0,y_0)}}^\eta }+\sum\limits_{l'\in \Gamma_{\mathcal{\tilde{S}}^{('PS)}_m}} \frac{\beta S_m'}{ {d_{l',(x_0,y_0)}}^\eta }+N_0 B_m}
\label{Eq_SINR_PowerScale_Local}
\end{multline}
For a user situated well within LSA1, the first term of the numerator contributes to the dominant signal power. However, the dominant interference term in the denominator is always the second term (the term with $l'\in \Gamma_{\mathcal{\tilde{S}}^{('PS)}_m}$) because it will be the closest interfering cell to the LSA. This ensures that the power scaling method improves the SINR of the local contents by reducing $\beta$ for $0\le \beta \le 1$.

Since total transmit power of $P_t = \sum_{m=1}^{M} S_m$ is available, the power freed up by scaling down $S_m$ to $\beta S_m$ can be used to boost the global content ($m=1$). Thus, the global content power within the buffer zone becomes $\tilde{S}_0 = P_t - \sum_{m=1}^{M} \beta S_m$. Thus, for $m=1$, the SINR expression for the global content, $\gamma_1$, is given by,
\begin{equation}
\gamma_1 = \frac{\sum\limits_{l\in \Gamma_{\mathcal{S}^{(PS)}_1}} \frac{S_1}{ {d^{\ \eta}_{l,(x_0,y_0)}} } + \sum\limits_{l\in \Gamma_{\mathcal{\tilde{S}}^{(PS)}_1}} \frac{P_t - \sum\limits_{m=1}^{M} \beta S_m}{ {d^{\ \eta}_{l,(x_0,y_0)}} }}{N_0 B_1}
\label{Eq_SINR_PowerScale_Global}
\end{equation}
As the global content has no interference, the SINR and SNR expressions are equivalent. It is evident from \eqref{Eq_SINR_PowerScale_Global} that as $\beta$ reduces, the SINR of the global content increases and so does its coverage. Thus, the power scaling method increases the SINR (and hence the coverage area) of both the local and the global contents.

\subsubsection*{(B2) Spectral efficiency of IM-LSI through power scaling}
Since IM-LSI through power scaling follows reuse-1 for all contents, the spectral efficiency expression reduces to

\begin{equation}
    \xi_{IM-LSI,PS} = \frac{\sum_{m=1}^M \left( \frac{|\mathcal{S}_{m}|\times log_2 \mu_m}{T_{sym}}\right)}{\sum_{m=1}^M B_m} \text{ bits/s/Hz}
    \label{SE_IMLSIPS}
\end{equation}

Note that power scaling does not affect spectral efficiency. However, simulation results show that it reduces local content coverage in the buffer zone.
\subsection{IM-LSI through orthogonality}
The O-LSI design for Fig. \ref{CellGrid} employed a frequency reuse factor of 2 for the local content throughout the service area. This higher reuse factor can be limited to the buffer zone. That is, the LSA excluding the buffer zone will employ reuse-1 for all contents while the buffer zone will use reuse-2. 
\subsubsection*{(C1) SINR for IM-LSI through orthogonality}
The SINR expression should account for the left and right buffer zones transmitting local contents in mutually exclusive sets of subcarriers.

\begin{equation}
\gamma_m = \frac{\sum\limits_{l\in \Gamma_{\mathcal{S}^{(IMO)}_m} \cup \Gamma_{\mathcal{\tilde{S}}^{(IMO)}_m}} \frac{S_m}{ {d_{l,(x_0,y_0)}}^\eta }}{ \sum\limits_{l'\in \Gamma_{\mathcal{S'}^{(IMO)}_m} \cup \Gamma_{\mathcal{\tilde{S}}^{'(IMO)}_m}} \frac{S_m'}{ {d_{l',(x_0,y_0)}}^\eta }+N_0 B_m}
\label{Eq_SINR_Ortho_Local}
\end{equation}

Observe that $\Gamma_{\mathcal{\tilde{S}}^{(IMO)}_m} = \emptyset$, which is the null set, when the $m^{th}$ content is not transmitted in the LB. Similarly, $\Gamma_{\mathcal{\tilde{S}}^{'(IMO)}_m} = \emptyset$ when the $m^{th}$ content is not transmitted in the RB. This design ensures that $\Gamma_{\mathcal{\tilde{S}}^{(IMO)}_m} = \emptyset \implies \Gamma_{\mathcal{\tilde{S}}^{'(IMO)}_m} \neq \emptyset$ and vice versa.
\vspace{0.15cm}
\subsubsection*{(C2) Spectral efficiency of IM-LSI through orthogonality}
The theoretical bit rate for local contents should be scaled by the fraction of cells that transmit that content in a particular local service area. Thus, the spectral efficiency expression becomes
\medmuskip=0mu
\thinmuskip=0mu
\thickmuskip=0mu
\begin{equation}
    \xi_{IM-LSI,O}\ =\  \frac{\sum_{m=1}^M \left( \frac{|\Gamma_{\mathcal{S}^{(IMO)}_m \cup \mathcal{\tilde{S}}^{(IMO)}_m }|}{|\Gamma_{\mathcal{S}^{(IMO)}_1 \cup \mathcal{\tilde{S}}^{(IMO)}_1}|} \frac{|\mathcal{S}_{m}|\times log_2 \mu_m}{T_{sym}}\right)}{\sum_{m=1}^M B_m} 
    \label{SE_IMLSI_O}
\end{equation}
\medmuskip=3mu
\thinmuskip=4mu
\thickmuskip=5mu
\section{Simulation Results}
The simulation study considers one global content and two local contents ($M=3$) that occupy equal bandwidths. Fig. \ref{CellGrid} shows two co-channel LSAs deployed next to each other. The simulation parameters are based on the system-level simulation assumptions described in \cite{3gppR12009308} and \cite{ibanez20215g} for a carrier frequency of 700 MHz. The Hata pathloss model, popular for pathloss predictions in the ultra-high frequency (UHF) band, is used here. Shadowing and small-scale fading are not modelled as their effects are assumed to be averaged out over the service area and over the bandwidth, respectively.
\subsection*{(i) User coverage}
\begin{figure}
    \centering
    \includegraphics[width=\columnwidth]{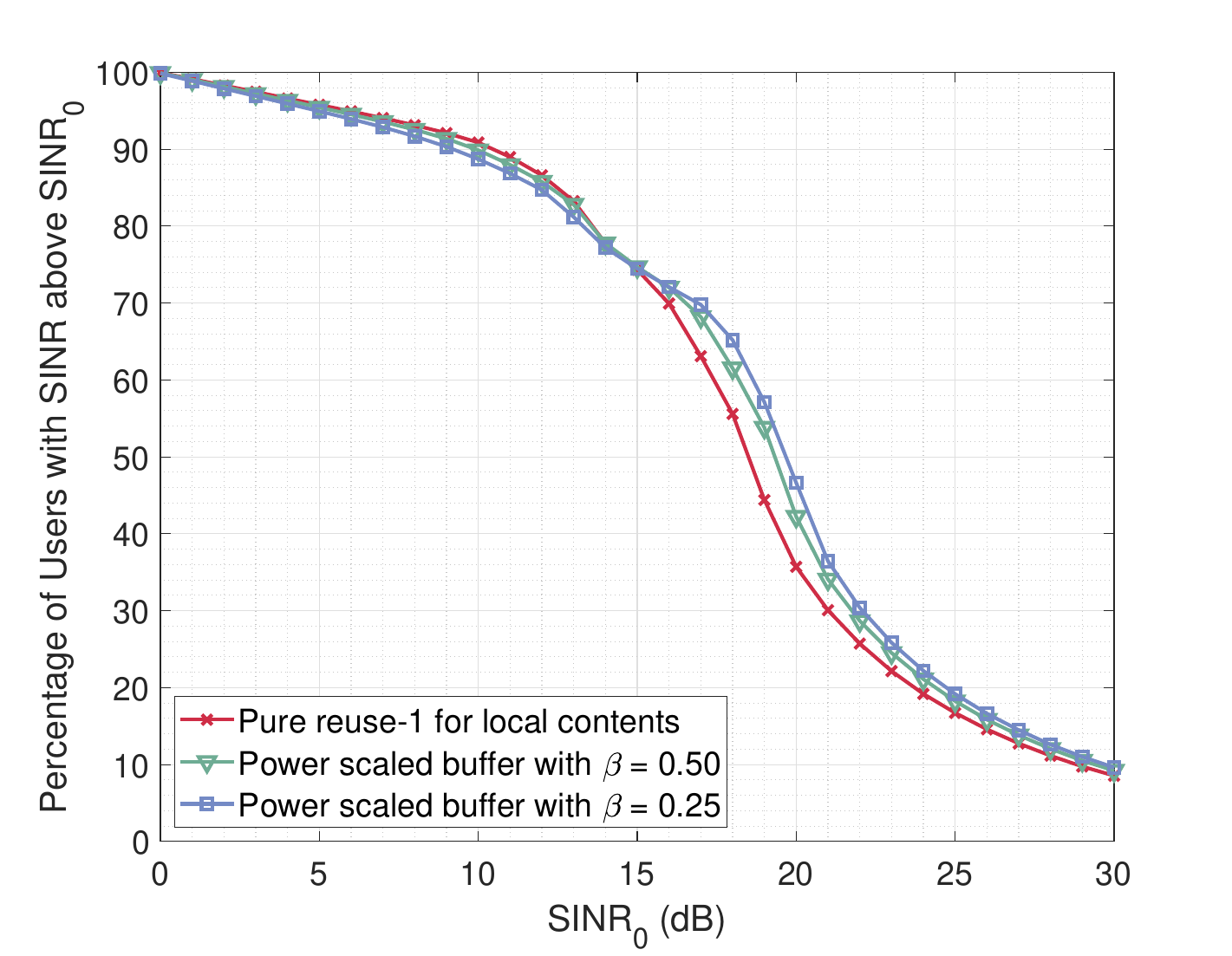}
    \caption{User coverage plot for IM-LSI through power scaling}
    \label{AreaPlot_scaling}
\end{figure}

The first set of simulations shall show the percentage of users covered by a particular scheme within area A1 (see Fig. \ref{CellGrid}), assuming that the users are uniformly distributed in the service area. A user is assumed to receive a signal of adequate quality if the SINR experienced by that user is greater than a threshold value $SINR_0$. This threshold is dependent on the modulation and code rate used. Fig. \ref{AreaPlot_scaling} shows this plot for IM-LSI through power scaling. A lower value of $\beta$ gives a coverage improvement of $11 \%$ at a threshold SINR of $20$dB. Thus, higher modulation orders are likely to see higher coverage with this scheme. As the O-LSI scheme is spectrally inefficient and since it allows only half as many local contents as the proposed scheme, it is not considered for this simulation.

\begin{figure}
    \centering
    \includegraphics[width=\columnwidth]{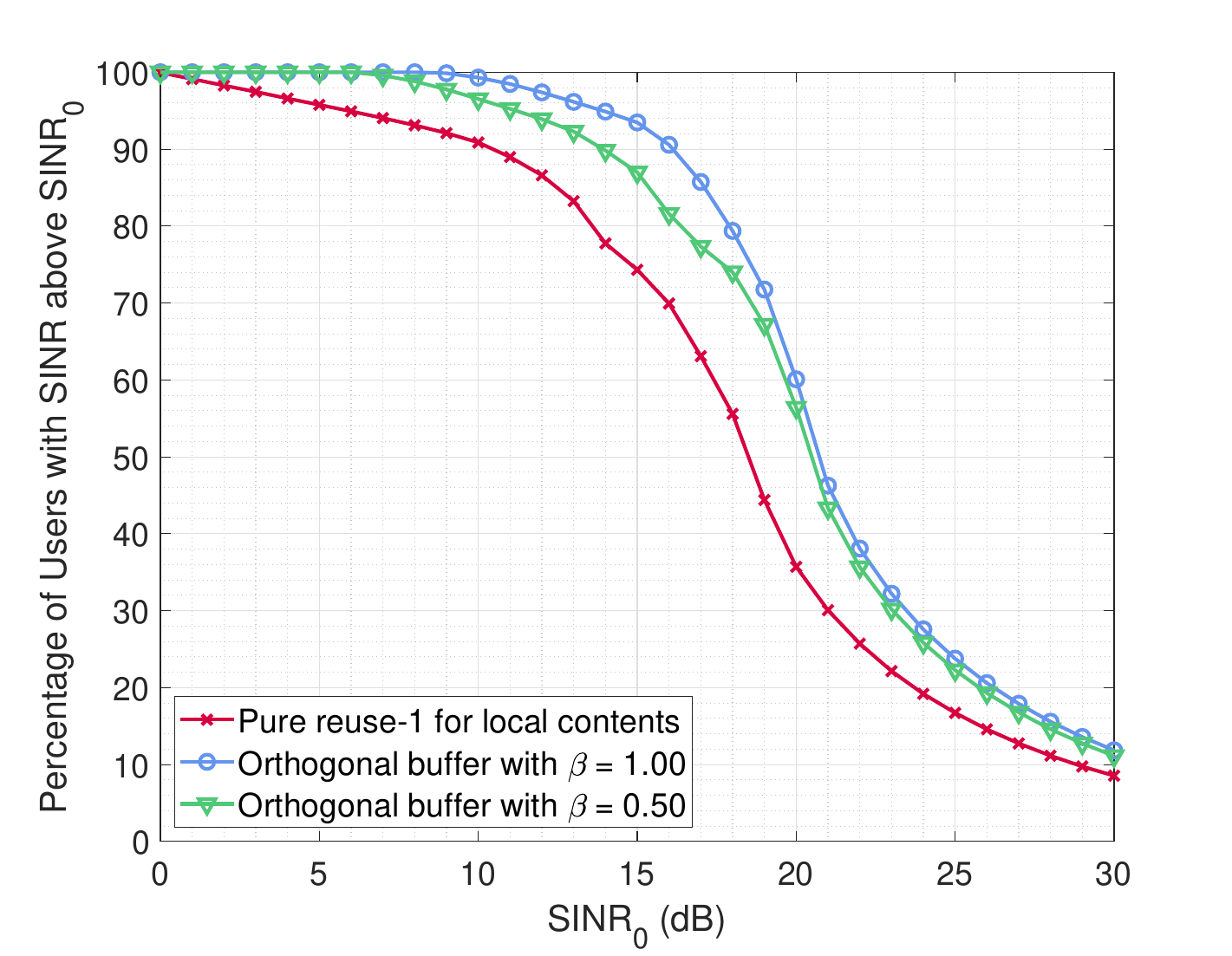}
    \caption{User coverage plot for IM-LSI through orthogonality (Content 2)}
    \label{AreaPlot_ortho}
\end{figure}

\begin{figure}[t]
    \centering
    \includegraphics[width=\columnwidth]{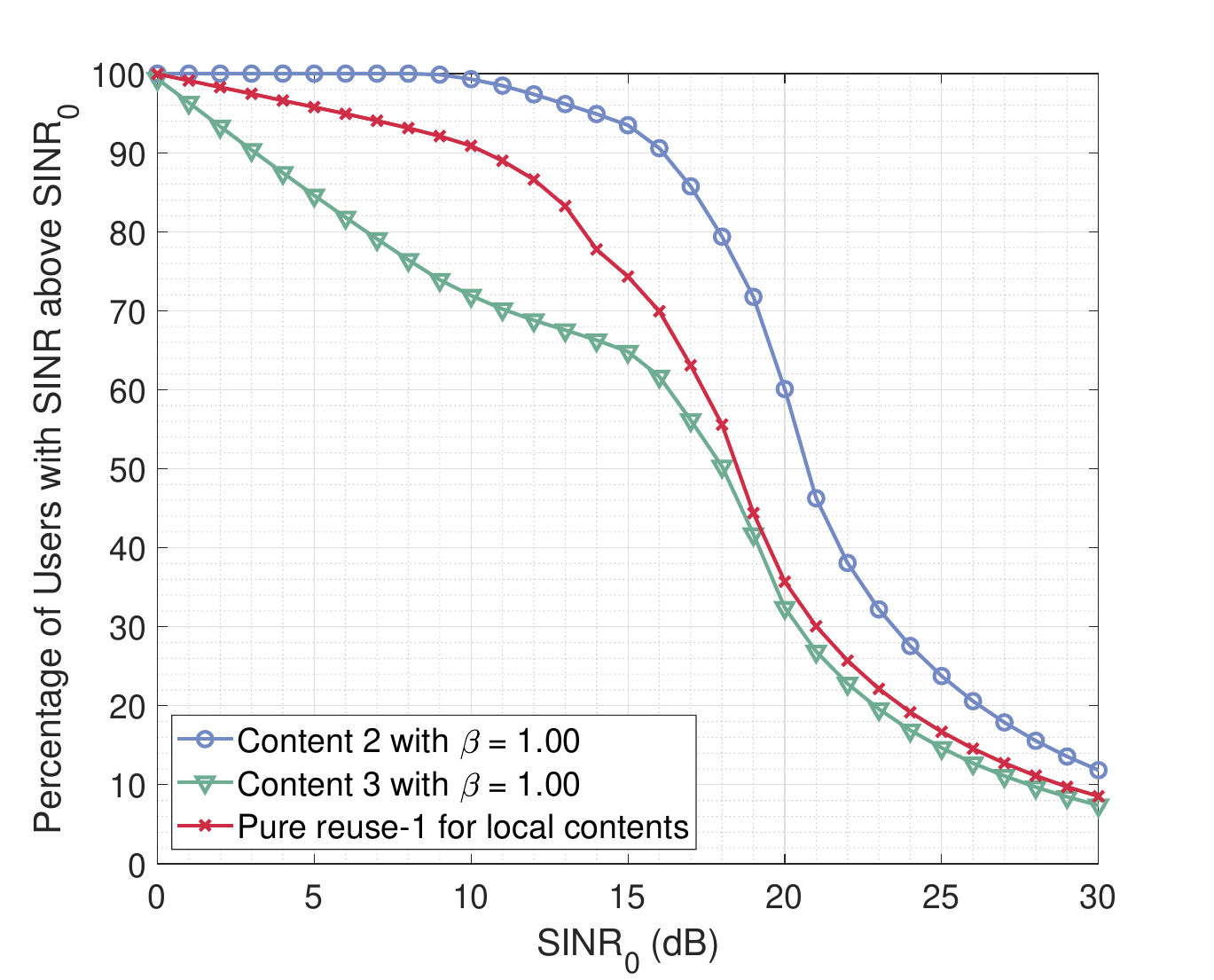}
    \caption{User coverage plot for IM-LSI through orthogonality \hbox{(Content 2 vs. 3)}}
    \label{AreaPlot_ortho2}
\end{figure}

Fig. \ref{AreaPlot_ortho} shows this plot for IM-LSI through orthogonality. A combination of orthogonality and scaled power ($\beta=0.5$) at the buffer zone is also tested. Scaling down power degrades this scheme's performance, so $\beta=1$ is recommended. We see a $19 \%$ coverage improvement at an SINR threshold of 15dB and a $24 \%$ improvement at an SINR threshold of 20dB. However, as the buffer uses an O-LSI-like deployment, half of the local contents do not receive this improvement. Fig. \ref{AreaPlot_ortho2} compares the coverage of content 2 and 3 in LSA1. As the buffer zone uses reuse-2 for local contents, the coverage of content 3 is significantly lower than content 2. However, power scaling approach provides equal coverage for all local contents. 


\begin{table} 
        \centering
        \captionsetup{justification=centering, labelsep=newline,font=sc}
        \caption{Comparison of coverage percentage for IM-LSI schemes}
        \begin{tabular}{l p{20mm} p{20mm}}
            \toprule
                & $SINR_0$ \newline $15\ dB$ & $SINR_0$ \newline $20\ dB$ \\
            \midrule 
             Content 2, IM-LSI-O & 93.5 & 60.1\\
             Content 3, IM-LSI-O & 64.8 & 32.4\\
             Average IM-LSI-O & 79.2 & 46.3\\
             IM-LSI-PS ($\beta=\frac{1}{4}$)  & 74.5 & 46.6\\
             Pure reuse-1 & 74.3 &35.7\\ 
            \bottomrule
        \end{tabular}
        \\ \hfill \\
\hfill  PS: Power Scaling \hfill O: Orthogonality\hfill 
        \label{Table0}
\end{table}

Table \ref{Table0} compares the coverage of IM-LSI schemes as discussed in the previous paragraphs. Since IM-LSI by orthogonality has differing coverage for local contents, its average shall be taken as the representative value. Both approaches have a similar advantage over the `no buffer' case for an SINR threshold of 20dB, but IM-LSI through orthogonality has an advantage at an SINR threshold of 15dB. 
\subsection*{(ii) Content coverage}
\begin{figure}[t]
    \centering
    \includegraphics[width=\columnwidth]{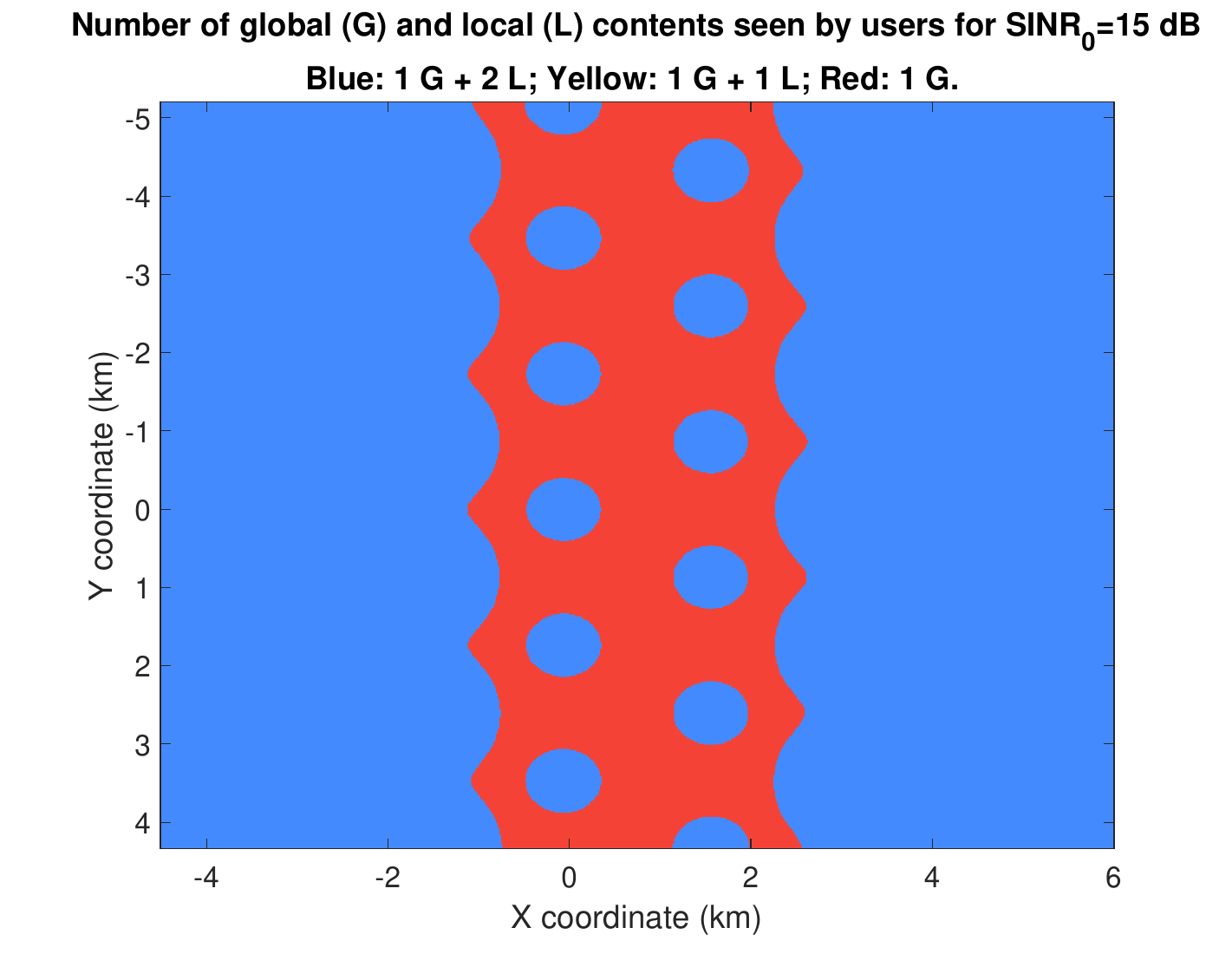}
    \caption{Number of contents seen by users for IM-LSI through power scaling}
    \label{Contentplot_scaling}
\end{figure}

\begin{figure}[t]
    \centering
    \includegraphics[width=\columnwidth]{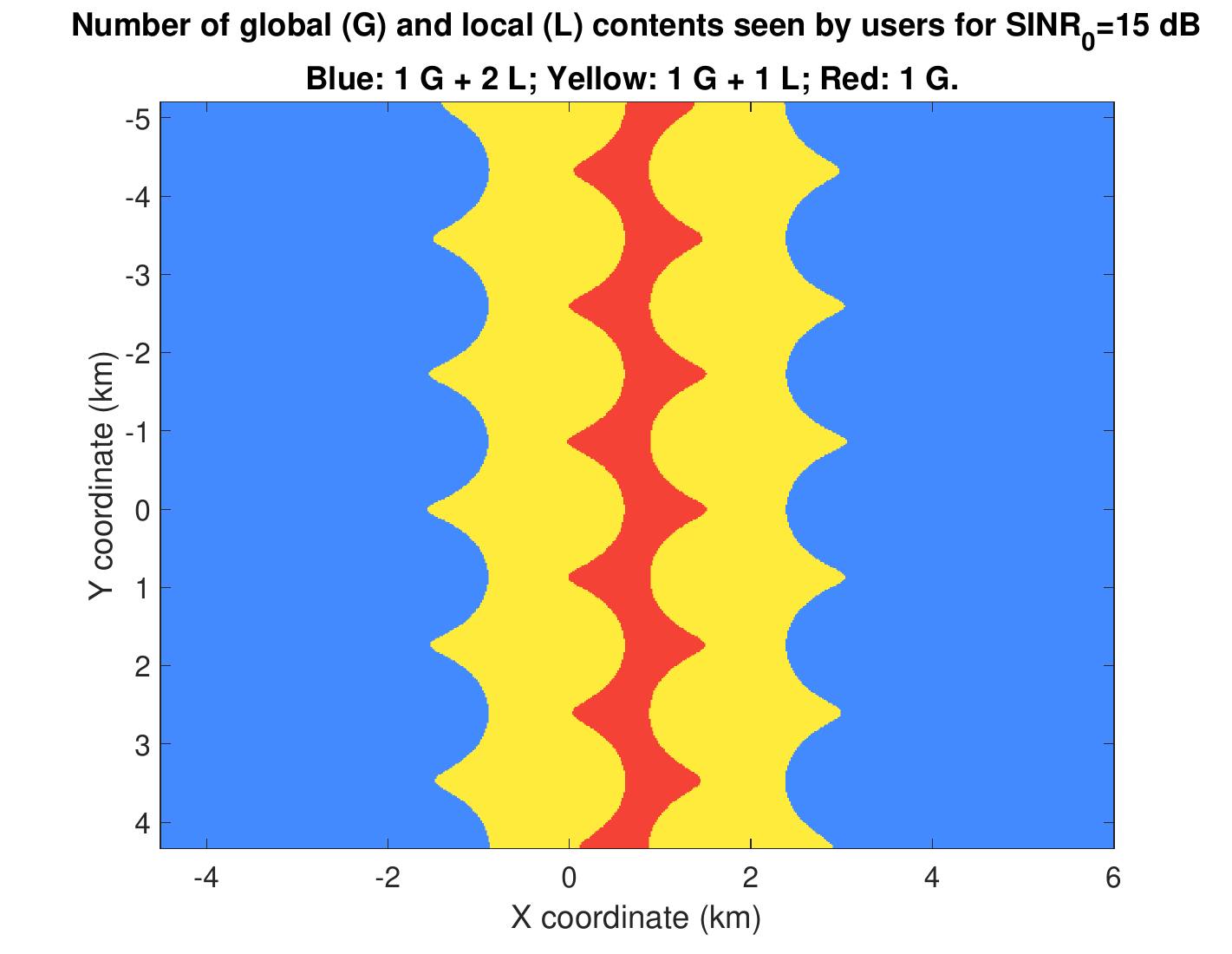}
    \caption{Number of contents seen by users for IM-LSI through orthogonality}
    \label{Contentplot_ortho}
\end{figure}

Next, we look at the number of contents seen by a receiver at different locations of area A2 (see Fig. \ref{CellGrid}), assuming that signals with SINR above the threshold can be decoded with satisfactory performance. Fig. \ref{Contentplot_scaling} and Fig. \ref{Contentplot_ortho} show this with the help of a color map. Irrespective of the scheme, the receivers see all the local contents outside the buffer zone with IM-LSI schemes. O-LSI, however, only allows half of the local contents in either LSA. The same plots for O-LSI would show two contents (one global and one local) throughout the area A2. The power scaling approach allows three contents near the transmitters at the buffer zone and only the global content in the remaining area of the buffer. The orthogonal approach allows at least one local content in most of the buffer zone and the global content alone at the immediate LSA boundary. 

We now quantify the plots in Fig. \ref{Contentplot_scaling} and Fig. \ref{Contentplot_ortho}. For an SINR threshold of 15dB, the orthogonal buffer allows all three contents at 65.5\% of the area A2. At least two contents can be received over 94.2\% of the area, and the global content can be received everywhere. The power-scaled buffer allows all three contents over 74.9\% of the area. No region receives just two contents, and the global content can be received everywhere as earlier. 
\subsection*{(iii) Quantitative comparison of spectral efficiency}
We assume the same modulation order and bandwidth allocation for all contents. The ratio of equations \eqref{SEOLSI} and \eqref{SE_IMLSIPS} reveals that O-LSI spectral efficiency is poorer by a factor of $\frac{1}{M}\sum_{m=1}^M \frac{|\Gamma_{\mathcal{S}^{(O)}_m \cup \mathcal{\tilde{S}}^{(O)}_m }|}{|\Gamma_{\mathcal{S}^{(O)}_1 \cup \mathcal{\tilde{S}}^{(O)}_1}|}=\frac{2}{3}$ for $M=3$ compared to IM-LSI through power scaling. This follows from the fact that a third of the subcarriers are nulled in each LSA to ensure orthogonality while no such nulling happens in the power scaling approach. For one global and $M-1$ local contents, O-LSI would be poorer by a factor of $\frac{1+(M-1)/2}{M}$.

The ratio of equations \eqref{SEOLSI} and \eqref{SE_IMLSI_O} shows a spectral efficiency degradation of $\sum_{m=1}^M \frac{|\Gamma_{\mathcal{S}^{(O)}_m \cup \mathcal{\tilde{S}}^{(O)}_m }|}{|\Gamma_{\mathcal{S}^{(IMO)}_m \cup \mathcal{\tilde{S}}^{(IMO)}_m }|}$ for O-LSI, since $\Gamma_{\mathcal{S}_1 \cup \mathcal{\tilde{S}}_1}$ is equal to the total cells in the LSA irrespective of the scheme used. For the system shown in Fig. \ref{CellGrid}, 8 cells of LSA1 out of 40 lies in the buffer zone. Thus, O-LSI spectral efficiency will be  $\frac{40+40+0}{40+40+32}=0.71$ times that of IM-LSI with orthogonality. 
Finally, the ratio of \eqref{SE_IMLSIPS} and \eqref{SE_IMLSI_O} is $M\div\sum_{m=1}^M\frac{|\Gamma_{\mathcal{S}^{(IMO)}_m \cup \mathcal{\tilde{S}}^{(IMO)}_m }|}{|\Gamma_{\mathcal{S}^{(IMO)}_1 \cup \mathcal{\tilde{S}}^{(IMO)}_1}|}=1.07$, which shows that the power scaling scheme has a marginally higher spectral efficiency than IM-LSI based on orthogonality. The exact numbers depend on the number of cells in the LSA and the buffer zone.
\section{Conclusion}
The proposed IM-LSI approaches have significant spectral efficiency gains compared to the O-LSI scheme \cite{Lopez2014Technical}. The number of local contents that can be served doubles for IM-LSI schemes although the coverage marginally degrades near the LSA boundary. Also, IM-LSI coverage area is higher by as much as $24 \%$ as compared to a pure reuse-1 deployment with no loss in spectral efficiency.

\bibliographystyle{IEEEtran}
\bibliography{IEEEabrv,mybibfile}

\vfill

\end{document}